# PERCEIVED ANNOYANCE IN MULTI-SOURCE ELECTRIC VEHICLE AVAS ENVIRONMENTS

Berkay KULLUKCU

*Chair of Acoustics and Haptics, TU Dresden, 01062 Dresden, Germany*

*email: berkay.kullukcu@tu-dresden.de*

Jonas KRAUTWURM

*Chair of Acoustics and Haptics, TU Dresden, 01062 Dresden, Germany*

*email: jonas_lukas.krautwurm@tu-dresden.de*

Serkan ATAMER

*Chair of Acoustics and Haptics, TU Dresden, 01062 Dresden, Germany*

*email: serkan.atamer@tu-dresden.de*

Ercan ALTINSOY

*Chair of Acoustics and Haptics, TU Dresden, 01062 Dresden, Germany*

*Centre for Tactile Internet with Human-in-the-Loop (CeTI), TU Dresden, Dresden, 01069, Saxony, Germany*

*Research Cluster 6G-life, TU Dresden, Dresden, 01069, Saxony, Germany*

*email: ercan.altinsoy@tu-dresden.de*

The increasing usage of electric vehicles in urban environments has resulted in a widespread presence of AVAS sounds. While individual vehicle sound design and testing is a common approach, real-world traffic scenarios often involve the simultaneous presence of multiple vehicles. Their combined presence may lead to changes in perception, compared to when they are presented individually, specifically regarding annoyance. The work addresses annoyance perception in scenarios involving multiple electric vehicle AVAS sounds. It changes the traditional isolated source-based view into a scene-based one by investigating the combined presence of multiple vehicle sounds as they are experienced in realistic traffic environments. Binaural listening tests were conducted using recorded electric vehicle pass-by sounds. The stimuli presented different traffic scenarios, including single and multiple-vehicles. Selected stimuli were spatially arranged to simulate vehicles approaching from opposite directions. After each stimulus, participants rated their perceived annoyance, enabling a comparison of annoyance responses between isolated and multiple AVAS sound scenarios. The test investigated how different levels of overlapping AVAS sounds affect perceived annoyance when multiple electric vehicles are passing by.

---

## 1. Introduction

The electrification of road transport is changing the soundscapes of our cities. At low speeds, the propulsion noise of electric and hybrid vehicles is often substantially reduced in comparison to conventional internal combustion engine vehicles, which is beneficial to the sound quality and public health in our cities. However, the reduced audibility has also been linked to safety concerns regarding visually impaired pedestrians who often rely on their ability to hear the approaching vehicle to understand the vehicle's presence, distance, and direction of approach [1-6].

Most of the evaluation strategies for AVAS detection, identification, and sound quality follow a source-centric approach. Early studies focused on quantifying the increase in crash risk for hybrid vehicles during low-speed manoeuvres and confirmed the detection of quiet vehicles at shorter distances than conventional vehicles [3, 4]. Other studies focused on quantifying reaction times or detection distances using binaural recording and controlled soundscapes, demonstrating strong dependencies on background noise, signal spectra, and temporal/spectral modulation effects [5, 6, 12–17]. These studies are all highly relevant but follow an implicit assumption that the target vehicle is auditorily separable and can be attributed to a single source.

From the environmental noise point of view, the interaction of two or more noise sources has been well understood and studied over the years. However, in the case of the long-term response of the community to environmental noise, it is not possible to predict the combined annoyance response through the simple addition of the energy from the two noise sources. This is why the annoyance equivalent model transforms the noise from the two sources into the equivalent exposure to a reference source and then adds them up [18]. Environmental noise standards acknowledge the co-occurrence of two or more noise sources and recommend the systematic description and reporting of the annoyance results [11, 19]. However, these are usually related to long-term exposure (hours to years), whereas the AVAS perception in the case of road traffic usually involves short bursts (seconds), often with strong contextual cues such as the ability to see the source, motion, and the spatial location of the source.

As regards the short-term assessment of the sound quality and annoyance of AVAS signals, psychoacoustic models may be considered as providing valuable additional descriptors. These may include the Zwicker loudness model (ISO 532), and related descriptors such as sharpness (DIN 45692), as well as tonality, roughness, and fluctuation strength, which may help to relate the sound features to the listener's perception [20-23]. Annoyance models using psychoacoustics combine several descriptors to predict annoyance levels more accurately than single descriptors such as sound pressure levels, in particular tonal and modulated sounds [24]. However, it is not clear whether these models, which are normally derived from single sounds, can account for the interaction effects observed when several AVAS signals are present in the scene.

The soundscape in the city is not determined only by the sound pressure levels but is also related to the perception of the scene in which the sounds are embedded. This leads to the assessment of the scene as perceived in the context, rather than assessing the individual sources [9]. This change in assessment is particularly relevant to the assessment of AVAS signals because the sound from one vehicle may be acceptable in isolation but become problematic in the context of several sources: several tonal sources may beat together, make tones more prominent, and increase the perception of crowding in the scene.

Moreover, AVAS is not the sole contributor to low-speed exterior noise. Instead, tire-road and auxiliary sounds dominate the scene with increasing speed. The relative contribution of AVAS depends on the vehicle and ambient environments [9, 25, 26]. Therefore, the scene-based method should treat AVAS as part of the traffic soundscape.

Studies on combined source annoyance have demonstrated that annoyance is not necessarily related to the A-weighted sound levels of the sources in question. The annoyance equivalent calculation is the sum of the annoyance equivalent levels of the sources in question [18]. Several reviews and conference reports have demonstrated that the exposure-annoyance relationship varies from source to source, which means that the results of the annoyance equivalent calculation may not be accurate [27].

This study examines the effect of overlapping AVAS sources on the annoyance of people when sounds are played in a controlled environment. The general idea behind this work is that a scene with multiple

AVAS sources isn't the sum of the individual AVAS sources, as the overlapping of these sources can actually increase the annoyance of the scene, or in some cases, reduce it by the interaction of the tones.

## 2. Method

A synthesized sound from [28] (S1) was used in the listening tests together with binaural pass-by recordings of electric vehicles with AVAS as the source material (Ford Kuga (S2) and Ford E-Transit (S3)). The spectrograms of the sounds are given in Figure 1 a, b, c respectively. The vehicle movement was modelled as a constant velocity pass-by at 20 km/h directly in front of the listener (Figure 2). For the two-vehicle scenes, the spatial arrangement of the vehicles was varied on a two-lane road with opposing directions, and the time offset was varied to create simultaneous crossing, +1 s, and +2 s of delay between the two vehicles (one vehicle passes 1 or 2 s before the other vehicle, directly in front of the listener). The auralization was conducted using a virtual acoustics renderer (TASCAR) [29] with a ground reflector and HRTF-based receiver model. Tire rolling noise was added to the TASCAR environment and played together with each AVAS sound, taken from a real-world recording of a Cupra Born driving at a stable 20 km/h (Figure 3).

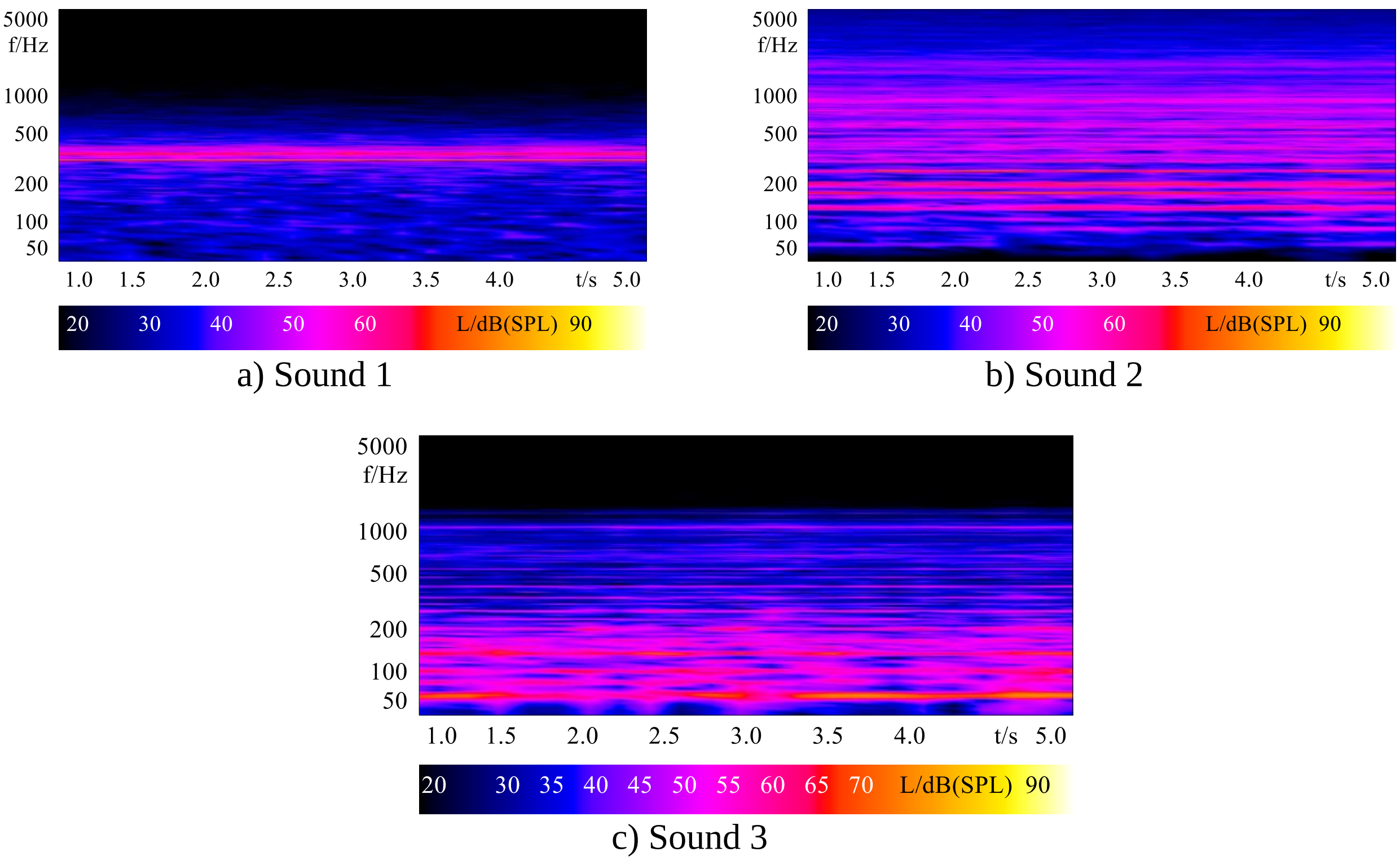

Figure 1: FFT vs time spectrograms of the sounds used in the experiment.

A total of N = 10 participants (5 males, 5 females, mean age: 30.7, std. dev.: 6.5 years) were presented with the stimuli through headphones in a quiet room and rated perceived annoyance on a continuous 0–100 scale for 13 stimuli (three single-vehicle pass-bys and ten two-vehicle scenes). Each stimulus was presented twice; ratings were averaged per participant and stimulus before group analysis. Head LabO2 was used as the sound reproduction device. All the AVAS sound files were maximum SPL equalized before they were used in the experiments.

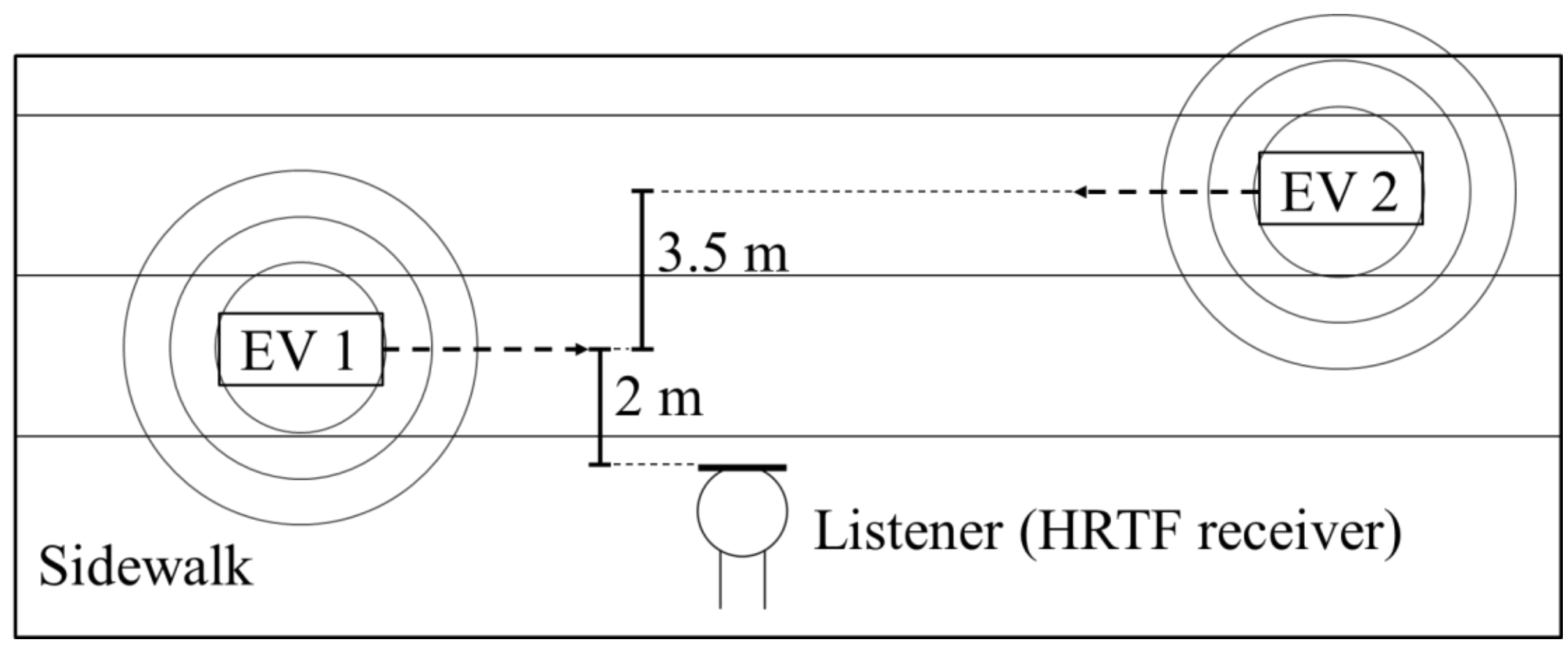

Figure 2: A 2D schematic of the simulation environment generated in TASCAR.

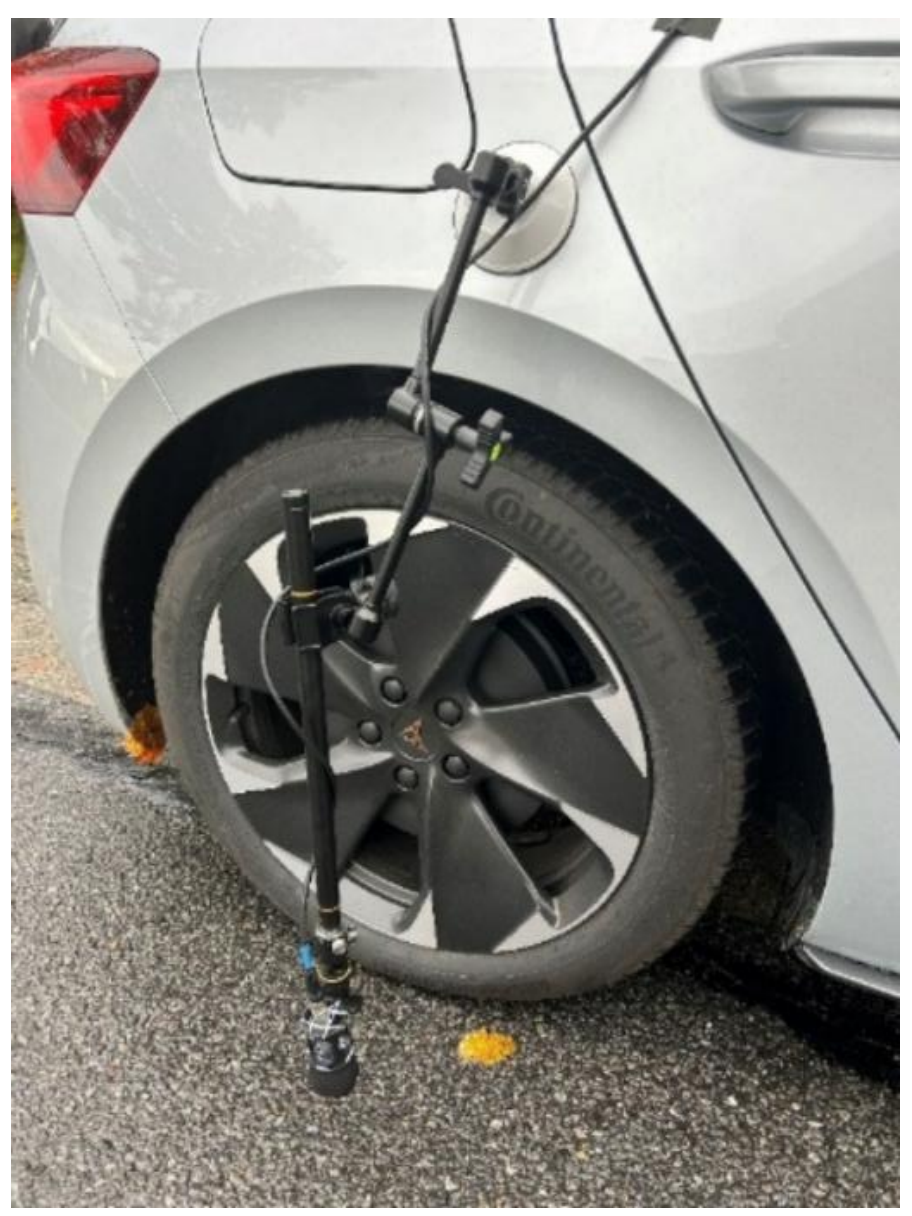

Figure 3: A microphone was placed directly towards to the rear-right wheel to decrease the interference of the vehicle`s own AVAS in the simulations

## 3. Results

Figure 4 summarizes raw annoyance ratings (mean across subjects ±95% CI) and their A-weighted sound pressure levels ($L_{Aeq}$), averaged for the 2 seconds where the cars were passing directly in front of the listener, for all stimuli. The list of names on the y-axis represents each stimulus, where the first sound name appears in multi-AVAS conditions is the AVAS of the vehicle coming from left to right (for example in S1-S3, S1 is the AVAS of the vehicle coming from left to right.)

Single-vehicle ratings were similar across sounds (means: 46.4 points when averaged across S1–S3; individual means: S1 (47.1±11.7), S2 (45.2±8.7), S3 (46.9±10.9)). Two-vehicle scenes showed higher average annoyance (means: simultaneous 52.9, time-offset 52.7). The highest mean annoyance occurred for S1+S3 (58.0±7.6), S1+S2 with 1 s offset (57.0±9.0), S1 with 2 s offset relative to S2 (56.7±9.1) (values in points ±95% CI). The mean within-subject increase was 6.54 points (95% CI [2.46, 10.63]). A paired t-test indicated a significant difference (t(9)=3.14, p=0.012), which was confirmed by a Wilcoxon signed-rank test (p=0.014). After Holm correction across three planned contrasts, the simultaneous-vs-single comparison remained significant (p=0.036). In addition to this, the overall psychoacoustic Impulsiveness (Hearing Model) showed a strong association with the mean annoyance (Spearman ρ = 0.823, p = 0.0006, BH–FDR q = 0.047).

Delayed (time-offset) two-vehicle scenes showed a similar mean increase (6.31 points), but with larger between-subject variability (95% CI [-1.32, 13.94]) and no significant difference from single-vehicle stimuli (p=0.139). When all two-vehicle scenes were pooled, the mean increase was 6.45 points (p=0.043), but this pooled test did not remain significant after Holm correction.

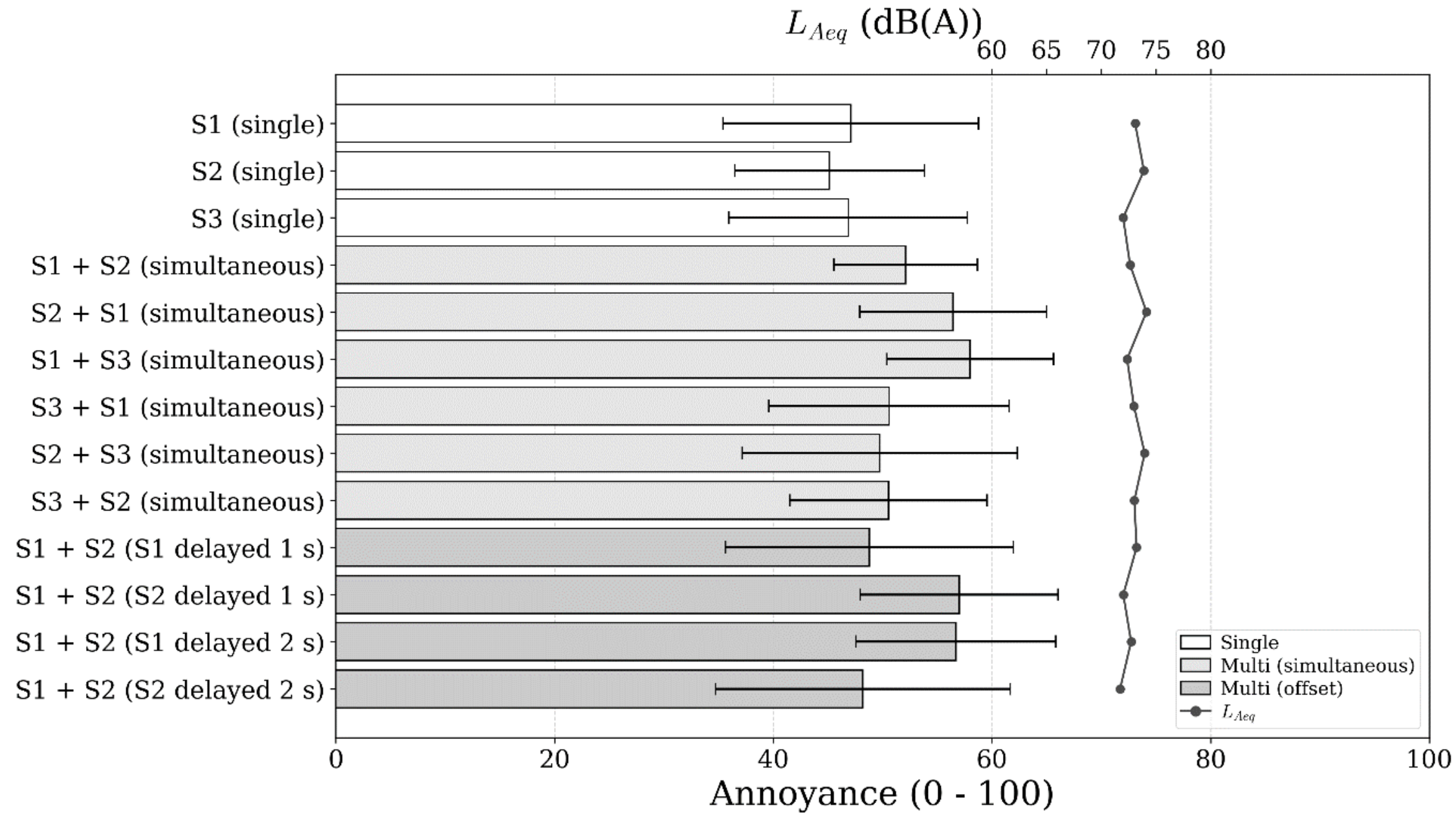

Figure 4: Results of the experiment, showing mean scores for all of the test conditions with the corresponding error bars. $L_{Aeq}$ ranges 71.73–74.13 dB, SD 0.77 dB (mean 72.91 dB).

We also tested whether swapping the approach direction/lane assignment changes annoyance by paired within-subject comparisons of stimuli with identical sound content but reversed order (S1+S2 vs S2+S1, S1+S3 vs S3+S1, S2+S3 vs S3+S2), also having the effect that the near-lane vehicle (first sound in the label) being acoustically louder at pass-by. No swap comparison was significant (S1/S2: p(t)=0.226, p(W)=0.264; S1/S3: p(t)=0.236, p(W)=0.264; S2/S3: p(t)=0.833, p(W)=0.846), and the combined direction index (participant-wise mean across the three swaps) was −0.75 points with 95% CI [−5.70, +4.20] (p(t)=0.773, p(W)=0.557). Additionally, $L_{Aeq}$ did not explain annoyance (Spearman ρ=-0.154, p=0.616).

## 4. Discussion

The results verify the underlying theory of short-term scene annoyance in multiple-source AVAS scenes being not possible to predict from single-vehicle results. Scenes with two vehicles with annoyance levels greater than the baseline, calculated as the average of their constituent single-vehicle AVAS, are depicted in Figure 4. The results are in line with the overall body of work on noise annoyance from multiple sources, where it is known that annoyance is non-additive and dependent on source interaction rather than simple summation of all noise sources. Although it is a short-term rather than a long-term study, with a corresponding short-term scene annoyance rather than a long-term community exposure, it is interesting to note that a similar effect is seen. The effect is dependent on the composition of the scene.

A possible explanation for this is the theory of auditory scene organization and information masking. For a scene with two moving vehicles, auditory organization is necessary. Recent work has highlighted that multiple electric vehicles can affect auditory localization and perceptual judgments in ways that are not captured by single-source testing [18]. Confusion can lead to a number of problems with annoyance, as it is possible that a greater annoyance is felt as a result of a confused brain attributing the sounds. This is in line with the theory of soundscape, where it is suggested that soundscape perception is of primary importance for evaluating a scene, rather than the sounds within it.

Lastly, the differences observed highlight the possible interest in integrating perceptual tests and psychoacoustic parameters and annoyance modelling [20-24]. A key psychoacoustic result is that overall Impulsiveness (Hearing Model) was the strongest correlate of mean annoyance.

## 5. Conclusion

This study aimed to explore the level of annoyance in short duration multi-source AVAS scenes using a binaural auralization technique with single- and two-vehicle pass-by sounds. In N=10 participants, simultaneous two-vehicle scenes were rated significantly more annoying than single-vehicle stimuli on the raw 0–100 scale. The result supports a scene-based view of AVAS evaluation, in which the multi-source context can change annoyance responses compared with isolated presentations. Furthermore, the increased annoyance in multi-source scenes could not be explained by residual level differences.

Limitations include the small number of participants, the limited set of stimuli, and the fixed speed. Future work is needed to increase the number of the stimulus set, including more AVAS scenes, types of vehicles, and types of background sounds. It is also important to extend the scene manipulation space to include more than two vehicles, varying the lane distance, and including realistic intersection scenarios. Finally, it is important to explore the use of psychoacoustic and scene-based metrics as predictors of multi-source AVAS annoyance, in order to allow the use of optimization models in relation to the design of AVAS scenes in realistic traffic scenarios.

## Acknowledgments

This work was supported by the Joint project 6G-life, under Grant 16KISK001K and as a part of Germany’s Excellence Strategy – EXC 2050/1 – Project ID 390696704 – Cluster of Excellence “Centre for Tactile Internet with Human-in-the-Loop” (CeTI) of TU Dresden. This work was supported by the Deutsche Forschungsgemeinschaft (DFG, German Research Foundation) -444809588 as a part of the priority program AUDICTIVE - SPP2236. This conference participation was supported by the German Academic Exchange Service (DAAD) under the Kongressreisenprogramm.

.